\documentclass[10pt,letterpaper,twocolumn]{article} 

\usepackage{ol2}
\usepackage[draft]{hyperref}
\usepackage{amsmath}

\begin{document}

\twocolumn[ 

\title{Optical modes in oxide-apertured micropillar cavities}

\author{Cristian Bonato$^1$, Jan Gudat$^1$, Keesjan de Vries$^1$, Susanna M. Thon$^{2,3}$, Hyochul Kim$^{2,4}$, Pierre M. Petroff$^2$, Martin P. van Exter$^1$, Dirk Bouwmeester$^{1,2}$}
\address{$^1$ Huygens Laboratory, Leiden University, P.O. Box 9504, 2300 RA Leiden, the Netherlands \\
$^2$ University of California Santa Barbara, Santa Barbara, California 93106, USA \\
$^3$ Current Address: University of Toronto, 10 King's College Road, Toronto, Ontario M5S 3G4, Canada \\
$^4$ Current Address: Department of ECE, IREAP, University of Maryland, College Park, Maryland 20742, USA
}

\begin{abstract}We present a detailed experimental characterization of the spectral and spatial structure of the confined optical modes for oxide-apertured micropillar cavities, showing good-quality Hermite-Gaussian profiles, easily mode-matched to
external fields. We further derive a relation between the frequency splitting of the transverse modes and the expected Purcell factor. Finally, we describe a technique to retrieve the profile of the confining refractive index distribution from the spatial
profiles of the modes.
\end{abstract}

] 
\maketitle
Microcavities embedding single self-assembled quantum dots (QDs) are promising devices for the implementation of single photon sources\cite{straufNP07} and hybrid quantum-information applications based on cavity quantum electrodynamics\cite{bonatoPRL2010}.
For such applications, a large Purcell factor ($P=\frac{3}{4\pi^2}\left(\frac{\lambda}{n}\right)^3\frac{Q}{V}$), is beneficial, i.e. the effective optical mode volume should be kept as small as possible, while maintaining a high quality factor, $Q$ \cite{vahalaNature03}.
Moreover, it is crucial for the cavity modes to be easily mode-matched to an external field, either for efficient light collection or to get high-contrast dipole-induced reflection\cite{waksPRL06, bonatoPRL2010}. \\
Etched micropillars with quality factors exceeding $150000$ have been demonstrated\cite{reitzensteinAPL07, forchelIOP10}, reaching the strong coupling regime\cite{reithmayerNature04}. The integration of electrical contacts in these structures, though possible, is not straightforward \cite{bocklerAPL08}.\\
In oxide-apertured micropillars, the transverse optical confinement is determined by the difference in the effective refractive index between the oxidized and un-oxidized regions \cite{zinoniAPL04, stoltzAPL05, ellisJPCM08}. This creates a gentle
confinement and reduces scattering compared to the rough side walls of etched micropillars. Moreover, structural contact to the bulk wafer is maintained, giving the ability to integrate electrical contacts in a convenient way
\cite{straufNP07, rakherPRL09}and enabling controlled electron charging of the QD and fine-tuning of the optical transition frequency via the quantum-confined Stark effect. Finally, for application involving electron spins, polarization-degeneracy of the
cavity modes is required. This can be achieved for oxide-apertured micropillars in a non-invasive way, controlling strain by laser-induced surface defects \cite{bonatoAPL2009, gudatAPL11} burned far-away from the cavity.\\
In this paper, we present a detailed experimental characterization of the spectral and spatial profiles of the optical modes for oxide-apertured micropillar cavities, showing that these structures exhibit high-quality Hermite-Gaussian modes, easy to
mode-match to external optical fields. Furthermore, we show that the higher-order optical modes can be used to gain detailed information about the structure. First of all, we derive a relation between the expected Purcell factor and the frequency splitting between the fundamental and the first excited mode, which can be used for quick estimate of the suitability of the device for cavity quantum electrodynamics. Finally, we introduce a technique to reconstruct the profile of the confining refractive index profile from the experimentally measured mode profiles. This technique can be helpful to characterize non-invasively the internal structure of a microcavity.

Micropillars are grown by molecular beam epitaxy on a GaAs [100] substrate. Two distributed Bragg reflector (DBR) mirrors consisting of alternating $\lambda/4$-thick layers of GaAs and Al$_{0.9}$Ga$_{0.1}$As embed a thin AlAs layer and the active layer with InGaAs/GaAs self-assembled QDs.
Trenches are etched down to the bottom DBR, leaving a circular pillar with a diameter of about $30$ $\mu$m (see Fig. 1a). The sample is steam-oxidized at $420^{\circ}$C, for a predetermined time based on calibration samples, to create an AlOx front in the AlAs layer, leaving an un-oxidized area with a controlled diameter between $3$ and $5$ $\mu$m. \\
\begin{figure}[htb]
\centerline{
\includegraphics[width=1\linewidth]{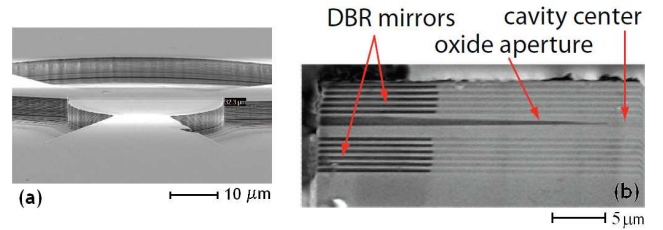}}
\caption{(a) SEM view of a micropillar cavity with three trenches. (b) SEM cross-section image of a typical micropillar. The AlOx layer has a linearly-decreasing thickness, leaving a small un-oxidized area at the center of the pillar.}
\label{fig:figure1}
\end{figure}
The optical modes are investigated by pumping the structure non-resonantly ($785$ nm, above the GaAs bandgap) with a few mW laser power, tightly focused on the sample by a microscope objective ($100$x, NA = $0.8$) to a spot of about $1$ $\mu$m. The
photoluminescence is collected with a multi-mode fiber and spectrally resolved with a spectrometer (resolution $0.016$ nm/pixel). In order to have a spatially-resolved photoluminescence plot, the excitation beam is scanned utilizing a piezo-driven xy-stage
at a step size of $0.1$ $\mu$m over an area of $10 \times 10$ $\mu$m. Such a setup, spatially scanning the pump beam and collecting the emitted photoluminescence with no spatial-selectivity, allows us to probe the mode intensities within the cavity layer, under the assumption of a spatially-constant distribution of emitters. This assumption is valid for our sample, given the high dot density and the pumping power.

\begin{figure}[htb]
\centerline{
\includegraphics[width=\linewidth]{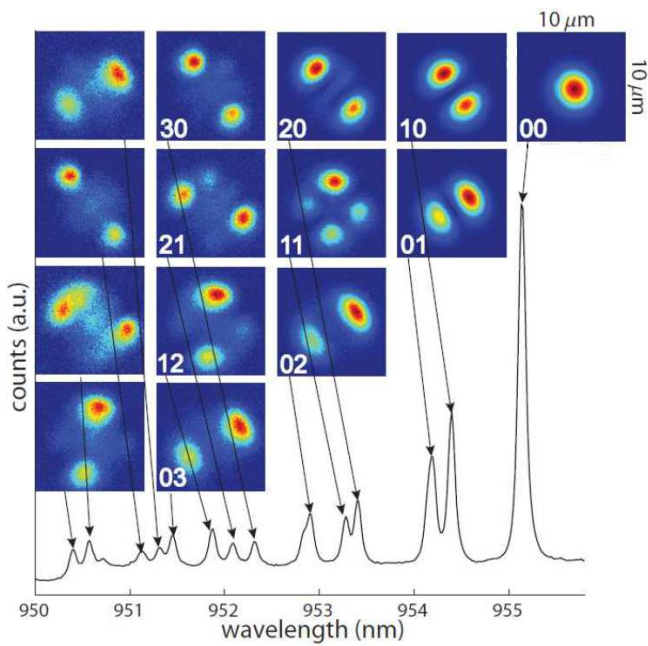}}
\caption
{Measured optical modes and spectrum of a non-polarization resolved spatial scan.}
\label{fig:optical_modes}
\end{figure}

A typical spectrum is presented in Fig.\ref{fig:optical_modes}. The spatial scans of the first ten peaks show clear features of Hermite-Gaussian modes and allow mode identification.
The fundamental cavity mode fits to a Gaussian profile with high accuracy, allowing easy mode-matching to external fields.
\\
In order to describe the optical modes we use an effective-index approximation\cite{hadleyOL95}, which simplifies the model by separating the longitudinal and transverse propagation and eliminating the longitudinal dependence by averaging it to an
effective transverse confinement. We model the oxidation taper as a quadratic refractive index profile\cite{hegblomAPL96}
\begin{equation}
n^{2}(x,y)=n_{0}^{2}\left(1-\frac{x^{2}}{r_{x}^{2}}-\frac{y^{2}}{r_{y}^{2}}\right).
\label{eq:neff}
\end{equation}
with $n_0$ the average refractive index of GaAs and Al$_{0.9}$Ga$_{0.1}$As.
The eigenfunctions of the associated Helmholtz equation, for the component of the wavefunction transverse to the propagation direction $z$, are
\begin{equation}
\psi_{[n,m]}(x,y)=H_{n} \left( \frac{\sqrt{2} x}{w_{x}} \right) H_{m} \left( \frac{\sqrt{2} y}{w_{y}} \right) e^{-\left( \frac{x^{2}}{w_{x}^{2}} + \frac{y^{2}}{w_{y}^{2}}\right)}
\label{eq:psi}
\end{equation}
where $H_n$ and $H_m$ are the Hermite-Gaussian functions of order $n$ and $m$, respectively. The waist of the fundamental mode is $w_{x,y}= \sqrt{r_{x,y} \lambda_0/ (\pi n)}$, where $\lambda_0$ is the vacuum wavelength. The eigenvalues associated with the
mentioned transverse modes, labeled as $[n,m]$, are the resonance wavelengths of the modes
\begin{equation}
\lambda_{[n,m]} \approx \lambda_{[00]} - n \Delta \lambda_x - m \Delta \lambda_y.
\label{eq:simplification}
\end{equation}
For a quadratic confining potential, the resonance wavelengths of the transverse modes are linearly spaced
\begin{equation}
\frac{\Delta \lambda_{x,y}}{\lambda_0} = \frac{1}{2} \left(\frac{\lambda_0}{n_0 \pi w_{x,y}}\right)^2.
\label{eq:transverse-mode-splitting}
\end{equation}
Gaussian fitting of the intensity profile of the fundamental mode shown in the measurement of Fig.\ref{fig:optical_modes} yields a spot size of $w_x\approx 2.13 \pm 0.08$ $\mu$m and $w_y\approx 2.25 \pm 0.09$ $\mu$m, fully compatible with widths
$w_x\approx 2.09$ $\mu$m and $w_y\approx 2.37$ $\mu$m calculated from the transverse mode splitting using Eq.\ref{eq:transverse-mode-splitting}.
From the mode waist one calculate the mode volume $V = (\pi/4) w_x w_y L_{eff}$. The effective cavity length $L_{eff}$ is given by $L_{eff} = L_{cav} +2 L_{pen}$, with $L_{pen} \approx \lambda_0/(4\Delta n)$ (penetration depth into DBR mirrors with index
contrast $\Delta n$). The Purcell factor can be rewritten as
\begin{equation}
P = \frac{12}{\pi} \mathcal{F} \left( \frac{\Delta \lambda}{\lambda}\right)
\end{equation}
where ($\mathcal{F} = \lambda_0 Q/ (2 n L_{eff}))$ is the cavity finesse.
This relation is important because it gives a quick way to characterize a microcavity, just looking at the spectral separation between the transverse modes. The larger the mode spacing, the smaller the cavity and the larger the Purcell factor.\\

\begin{figure}[htb]
\centerline{
\includegraphics[width=1\linewidth]{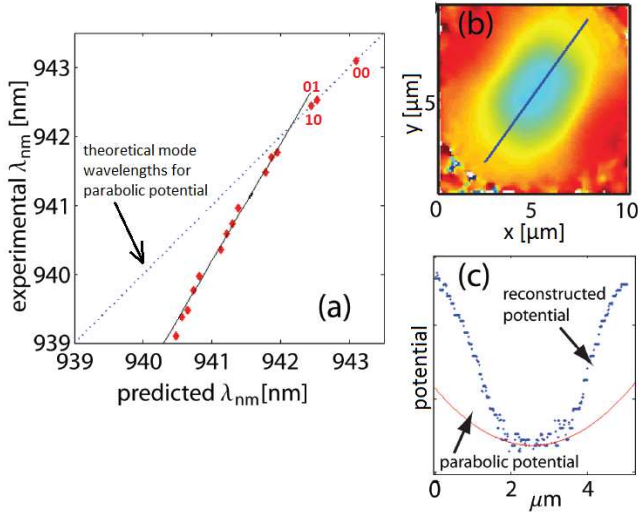}}
\caption{
(a)Super-linear mode frequency distribution of the experimental modes.
(b)Reconstruction of the refractive index profile.
(c)Refractive index profile cross-section taken from (b). The bottom is fitted to a parabolic potential.
}
\label{fig:potential}
\end{figure}

The quadratic model for $n^2 (x,y)$ only works for the lowest-order modes but does not correctly describe the resonance wavelengths of the higher-order modes. This is demonstrated in Fig. \ref{fig:potential}(a), which shows the measured mode
wavelength vs. the theoretical mode wavelength calculated with the parabolic refractive index profile in Eq.\ref{eq:neff} (for a different cavity than the one on Fig. 2). The experimental mode frequency distribution appears to be super-linear, suggesting
that the refractive index profile might be steeper than quadratic.
In order to determine the precise refractive index profile, we use the additional information contained in the intensity profiles of the eigenmodes. For this, we write the Helmoltz equation as $\mathcal{H} |\psi \rangle = 0$, where $\mathcal{H} =
\partial^2/\partial x^2 + \partial^2/\partial y^2 + V(x, y)$ and $V(x, y) = k_0^2 n^2 (x, y)$. Expressing $\mathcal{H}$ in terms of its complete set of eigenvalues $\varepsilon_n$ and orthonormal eigenfunctions $|\psi_n \rangle$, we get: $\mathcal{H} =
\sum_{n=1}^{\infty} \varepsilon_n \left| \psi_n \rangle \langle \psi_n \right|$. In $x$-representation (given that $\langle x \left| \partial^2/\partial x^2 \right| x \rangle = 0$),
\begin {equation}
V(x) = \langle x \left| \mathcal{H} \right| x \rangle =  \sum_{n=1}^{\infty} \varepsilon_n |\psi_n (x)|^2.
\end{equation}
The results, shown in Fig. \ref{fig:potential}b-c confirm that the refractive index profile is not quadratic, but consists of a flat
bottom that corresponds to the un-oxidized region surrounded by steeper walls resulting from the end of the oxidation front.

Here, we have only considered a scalar theory, justified by the fact that the cavity in Fig. 2 exhibits almost perfect polarization-degeneracy. In general, however, the modes show polarization properties. In particular, the fundamental mode is typically
split in two linearly-polarized submodes, with a frequency splitting ranging between $5$-$50$ GHz (compared to a linewidth of about $10$ GHz). Such polarization splitting can be ascribed to an asymmetry of the oxidation aperture given by different
oxidation rates along different crystal axes \cite{hegblomAPL96, coldrenIEEE97} or to an intrinsic birefringence of the material (possibly due to strain introduced by the etched trenches\cite{bonatoPRB11}).

In conclusion, we experimentally characterized spectrally and spatially the confined optical modes of oxide-apertured micropillars. The lower-order modes are well described by Hermite-Gaussian modes, which allows easy mode-matching to external fields. We
derived a formula to estimate the waist of the fundamental mode and the theoretical maximum Purcell factor from the splitting between the fundamental and the first-order modes. This equation can be used to quickly assess the quality of a device for
cavity-QED experiments. Finally, we presented a technique to retrieve the shape of the confining refractive index distribution determined by the oxidation layer, showing that it is flat in the center with steep walls, resulting in a super-linear
distribution of the confined modes.

This work was supported by the NSF grant 0901886, the Marie-Curie Award No. EXT-CT-2006-042580, and FOM/NWO Grant No. 09PR2721-2. We thank Nick G. Stoltz for the SEM image in Fig. 1b and Andor for the CCD camera.

\end{document}